\documentclass[showpacs,twocolumn,showkeys,amsmath,amssymb,pre]{revtex4-1}

\usepackage{bm}
\usepackage{bbm}
\usepackage{color}	
\usepackage{graphicx}   
\newcommand{\rd}{{\mathrm d}}
\newcommand{\re}{{\mathrm e}}
\newcommand{\ri}{{\mathrm i}}
\newcommand{\lF}{\langle\!\langle} 
\newcommand{\rF}{\rangle\!\rangle}                 

\begin{document}

\title[Variational principle]
      {Variational principle for time-periodic quantum systems} 

\author{Nils Kr\"uger}
	
\affiliation{Carl von Ossietzky Universit\"at, Institut f\"ur Physik,
	D-26111 Oldenburg, Germany}

\date{July 27, 2020}

\begin{abstract}
A variational principle enabling one to compute individual Floquet states 
of a periodically time-dependent quantum system is formulated, and successfully
tested against the benchmark system provided by the analytically solvable model
of a linearly driven harmonic oscillator. The principle is particularly well 
suited for tracing individual Floquet states through parameter space, and may 
allow one to obtain Floquet states even for very high-dimensional systems which
cannot be treated by the known standard numerical methods.
\end{abstract} 

	
\maketitle	


\section{Introduction}
\label{S_1}

The Rayleigh-Ritz variational principle has proven to be of outstanding 
practical value for the approximate determination of a quantum system's
ground state. For any trial state~$|\psi\rangle$ one has the inequality
\begin{equation}
	R \big[ |\psi\rangle \big] \equiv 
	\frac{\langle\psi | H | \psi\rangle}{\langle\psi|\psi\rangle}
	\ge E_0 \; ,
\label{eq:RRP}
\end{equation}	
where $H$ denotes the system's Hamiltonian, assumed to be time-independent,
and $E_0$ is its ground-state
energy. Hence, inserting an appropriate ansatz for the ground state, which 
should embody its key features on the one hand, and depend on convenient
variational parameters on the other, one obtains an upper bound on~$E_0$; 
mini\-mizing~$R$ with respect to the parameters, this bound often is found 
to be fairly tight. As has been concluded by Griffiths, ``the variational 
principle is extaordinarily powerful, and embarrassingly easy to 
use''~\cite{Griffiths05}.
  
The purpose of the present paper is to point out that there also is a 
variational principle which enables one to compute Floquet states of a 
periodically time-dependent quantum system. Such Floquet states have met 
with considerable interest recently; among others, they have been invoked 
for investigating the dynamics of atomic quantum gases in periodically driven 
optical lattices~\cite{Eckardt17}, and they lead to a natural explanation of 
the spontaneous breaking of time-translation symmetry occurring in so-called
Floquet time cystals~\cite{HolthausFlatte94,ElseEtAl16,vonKeyserlingkEtAl16,
YaoEtAl17}. Since their quasienergies constitute infinite, ladder-like classes 
of equally spaced representatives, the Floquet states do generally not possess 
a natural order; in particular, in most cases there is no ``Floquet ground 
state''. Thus, the idea of the Rayleigh-Ritz principle~(\ref{eq:RRP}) cannot 
be transferred one-by-one to Floquet systems. Nonetheless, it will be shown
that there exists a similar variational principle which may allow one 
to compute Floquet states even for systems which are so large that they are
no longer amenable to any other technique available so far.     

The paper is organized as follows: For the convenience of the reader, the
salient features of the Floquet picture are summarized in the subsequent
Sec.~\ref{S_2}. The new variational principle for Floquet states then is
formulated in Sec.~\ref{S_3}, and tested numerically with the help of an 
ana\-lytically solvable model system in Sec.~\ref{S_4}. An outlook towards
future applications of the principle is given in the concluding Sec.~\ref{S_5}.

\section{The Floquet concept}
\label{S_2}

Consider a quantum system governed by a Hamiltonian which depends periodically
on time~$t$ with period~$T$,   
\begin{equation}
	H(t) = H(t + T) \; ,
\end{equation}
acting on the system's Hilbert space~${\mathcal H}$. The Floquet theorem 
asserts that the associated time-evolution operator $U(t,0)$, mapping any 
initial state $| \psi(0) \rangle$ to the state $| \psi(t) \rangle$ which 
evolves from the former after time~$t$,
\begin{equation}
	| \psi(t) \rangle = U(t,0) \, | \psi(0) \rangle \; ,
\label{eq:TEP}
\end{equation}
can be factorized to read~\cite{Salzman74,BaroneEtAl77,GesztesyMitter81,
Holthaus16}
\begin{equation}
	U(t,0) = P(t) \exp(-\ri G t/\hbar) \; ,		  
\label{eq:UPG}	
\end{equation}
where $P(t) = P(t+T)$ is periodic in time and unitary, while the 
time-independent operator~$G$ is self-adjoint. With 
$P(0) = P(T) = {\mathbbm 1}$, the one-cycle evolution operator then
takes the form
\begin{equation}
	U(T,0) = \exp(-\ri G T/\hbar) \; .
\end{equation}
Generally, the spectral problem posed by $U(T,0)$ on ${\mathcal H}$ may be 
quite involved~\cite{Howland89a,Howland89b,Howland98,Howland92}; here we 
simply {\em assume\/} that $U(T,0)$ possesses a pure point spectrum of 
eigenvalues $\exp(-\ri\varepsilon_n T/\hbar)$ accompanied by normalized 
eigenstates~$| n \rangle$,
\begin{equation}
	U(T,0) \, | n \rangle = 
	\exp(-\ri\varepsilon_n T/\hbar) \, | n \rangle \; .
\label{eq:SEP}
\end{equation}	
This is not a trivial proposition; of course, this is always the case if
${\mathcal H}$ is of finite dimension. The so-called quasienergies 
$\varepsilon_n$ then constitute the eigenvalues of~$G$. Due to the 
multi-valuedness of the complex logarithm they are defined only up to an 
integer multiple of $\hbar\omega$, where $\omega = 2\pi/T$ is the angular 
frequency implied by the period~$T$.

The ``stroboscopic'' eigenvalue problem~(\ref{eq:SEP}) already leads to 
one of the decisive benefits of the Floquet picture. Expanding an initial
state with respect to the eigenstates $| n \rangle$ of $U(T,0)$,
\begin{equation}
	| \psi(0) \rangle  = \sum_n 
	| n \rangle \langle n | \psi(0) \rangle \; ,
\end{equation}
the combination of Eqs.~(\ref{eq:TEP}) and~(\ref{eq:UPG}) gives, for any 
time~$t$,
\begin{eqnarray}
	| \psi(t) \rangle & = &
	\sum_n \langle n | \psi(0) \rangle \, P(t) \exp(-\ri G t/\hbar) \,
	| n \rangle
\nonumber \\	& = &
	\sum_n \langle n | \psi(0) \rangle \, | u_n(t) \rangle
	\exp(-\ri\varepsilon_n t/\hbar) \; , 		
\label{eq:EVO}
\end{eqnarray}	  	  
where the {\em Floquet functions} 
\begin{equation}
	| u_n(t) \rangle = P(t) | n \rangle
\end{equation}
inherit the $T$-periodicity of $H(t)$ and $P(t)$, so that
\begin{equation}
	| u_n(t) \rangle = | u_n(t+T) \rangle \; . 	 		
\label{eq:PBC}
\end{equation}
Thus, when expanded with respect to the {\em Floquet states\/}
\begin{equation}
	| \psi_n(t) \rangle = | u_n(t) \rangle
	\exp(-\ri\varepsilon_n t/\hbar) \; ,
\label{eq:FLS}
\end{equation}
the time evolution~(\ref{eq:EVO}) of $|\psi(t)\rangle$ proceeds with 
{\em constant\/} amplitudes $ \langle n | \psi(0) \rangle$. In other words, 
the Floquet states carry time-independent occupation probabilities, despite 
the periodic time-dependence of their Hamiltonian, drastically simplifying 
the determination of the system's long-time behavior.

While the above stroboscopic approach is often found useful for numerical
purposes, there also exists another, ``extended'' viewpoint which is 
particularly helpful for conceptual considerations. Inserting a Floquet 
state~(\ref{eq:FLS}) into the time-dependent Schr\"odinger equation
\begin{equation}
	\ri\hbar\frac{\rd}{\rd t} | \psi(t) \rangle = 
	H(t) | \psi(t) \rangle \; ,
\end{equation}
one immediately finds
\begin{equation}
	\left( H(t) + \frac{\hbar}{\ri} \frac{\rd}{\rd t} \right)
	| u_n(t) \rangle = \varepsilon_n | u_n(t) \rangle \; .
\label{eq:EVP}
\end{equation}
Augmented by the periodic boundary condition~(\ref{eq:PBC}) to be satisfied
by the Floquet functions, this is an eigenvalue problem which does not pose
itself on the system's actual Hilbert space~${\mathcal H}$, but instead on an 
extended Hilbert space consisting of $T$-periodic functions~\cite{Sambe73}; 
this extended space is often denoted by $L_2[0,T] \otimes {\mathcal H}$ in 
the mathematical literature. Here the time~$t$ is no longer regarded as an 
evolution parameter in the sense of Eq.~(\ref{eq:TEP}), but rather as an 
additional coordinate; hence, the scalar product in this extended space
is given by
\begin{equation}
	\lF u | v \rF = \frac{1}{T} \int_0^T \! \rd t \,
	\langle u(t) | v(t) \rangle \; ,
\label{eq:SCP}
\end{equation}		     		 	
with $\langle \, \cdot \, | \, \cdot \, \rangle$ indicating the given scalar 
product on ${\mathcal H}$. Although it may seem strange from the viewpoint
of conventional quantum physics on ${\mathcal H}$, the operator
\begin{equation}
	p_t = \frac{\hbar}{\ri} \frac{\rd}{\rd t}
\end{equation}
acting on $L_2[0,T] \otimes {\mathcal H}$ now represents the momentum operator 
which is canonically conjugate to the coordinate~$t$; the periodic boundary 
condition~(\ref{eq:PBC}) makes sure that this operator is Hermitian. Observe 
that the quasienergy operator
\begin{equation}
	K = H(t) + p_t
\label{eq:KAM}
\end{equation}
which appears on the left-hand side of Eq.~(\ref{eq:EVP}) depends only linearly
on this momentum. As a consequence, its quasienergy spectrum is unbounded both 
from above and from below: Assume that $| u_n(t) \rangle \equiv 
| u_{n,0}(t) \rangle$ is a solution to the eigenvalue problem~(\ref{eq:EVP}), 
so that
\begin{equation}
	K | u_n (t) \rangle = \varepsilon_n | u_n (t) \rangle \; .
\label{eq:QEP}
\end{equation}	  	   	
Then for any integer $m = \pm 1,\pm2,\pm3,\ldots$ the functions 
\begin{equation}
	| u_{n,m} (t) \rangle = | u_n (t) \re^{\ri m \omega t}\rangle
\end{equation}
likewise are $T$-periodic eigensolutions,
\begin{equation}
	K | u_{n,m} (t) \rangle = 
	(\varepsilon_n + m \hbar\omega) | u_{n,m} (t) \rangle \; .
\end{equation} 	
Thus, from the perspective of the extended Hilbert space each Floquet 
state~(\ref{eq:FLS}) evolving in ${\mathcal H}$ is associated with 
infinitely many eigensolutions of the eigenvalue problem~(\ref{eq:EVP}) in
$L_2[0,T] \otimes {\mathcal H}$,
\begin{eqnarray} 
	& &	
	| u_n(t) \rangle \exp(-\ri\varepsilon_n t/\hbar) 
\nonumber \\	& = &	
	| u_{n,m}(t) \rangle 
	\exp\!\big[-\ri(\varepsilon_n + m \hbar\omega)/\hbar \big] \; . 
\end{eqnarray}
Therefore, within this extended approach a quasienergy should not be regarded 
as a number~$\varepsilon_n$, but rather as an infinite class of representatives
spaced by $\hbar\omega$,
\begin{equation}
	\varepsilon_n \equiv \{ \varepsilon_n + m \hbar\omega  \; | \;
	m = 0, \pm 1, \pm 2, \ldots \} \; ,
\label{eq:ILQ}
\end{equation}
reflecting the ``$\hbar\omega$-indeterminacy'' of the quasienergies
which stems from taking the complex logarithm of the Floquet multipliers
$\exp(-\ri\varepsilon_n T/\hbar)$ encountered in the stroboscopic approach.   

The observation that the quasienergy eigenvalue problem~(\ref{eq:QEP}) plays 
a role which is conceptually similar to that of the stationary Schr\"odinger 
equation now allows one to transfer many notions known from time-independent
quantum mechanics to periodically time-dependent quantum systems, such as
the Hellmann-Feynman theorem, or Rayleigh-Schr\"odinger perturbation 
theory~\cite{Sambe73}. There is, however, a notable exception: The fact that 
each Floquet state is equipped with an infinite ladder~(\ref{eq:ILQ}) of
quasienergies implies that these states cannot be ordered with respect to
the magnitude of their quasienergies, and there is no ``lowest'' quasienergy.
This means that the Rayleigh-Ritz principle~(\ref{eq:RRP}) has no immediate
counterpart in the extended Hilbert space, apparently depriving one of an 
efficient computational tool. In the follwing section it will be shown how 
this deficiency can be cured.

\section{Variational principle}
\label{S_3}

While the spectrum of the quasienergy operator~(\ref{eq:KAM}) is unbounded
from below, that of its square~$K^2$ is non-negative, as is the spectrum of 
$(K - \varepsilon)^2$ for any $\varepsilon$, be it an actual quasienergy 
eigenvalue of the system under consideration or not. Hence, one has the 
variational inequality  
\begin{equation}
	F_\varepsilon \big[ | \Psi \rangle \big] \equiv 
	\frac{ \lF \Psi | ( K - \varepsilon)^2 | \Psi \rF }
	{\lF \Psi | \Psi \rF } \ge 0 \; ,
\label{eq:FVP}
\end{equation}
where $| \Psi \rangle \in L_2[0,T] \otimes {\mathcal H}$ is a suitably 
parametrized $T$-periodic trial function, and double angular brackets indicate
the scalar product~(\ref{eq:SCP}). Evidently, this functional $F_{\varepsilon}$
adopts its minimum value zero if and only if $| \Psi \rangle$ indeed is an 
eigenfunction of $K$ with quasienergy eigenvalue~$\varepsilon$. Therefore, the 
inequality~(\ref{eq:FVP}) can be exploited in two substantially different ways:
{\em (i)\/} Keeping a given value of $\varepsilon$ fixed, and varying
$| \Psi \rangle$, one may investigate whether there exists a Floquet function 
with that particular quasienergy. {\em (ii)} A potentially more powerful 
application of the inequality~(\ref{eq:FVP}) emerges when $\varepsilon$ is
regarded as an additional variational parameter: In that case one may 
``follow'' an individual Floquet state in response to small changes of the
system's parameters, as is exemplified in the following section.  
 
In order to interpret the physical meaning of the numerical values adopted by 
the functional $F_\varepsilon$, consider an initial state in ${\mathcal H}$
at $t = t_0$,
\begin{equation}
	| \psi(t_0) \rangle = \sum_n a_n | u_n(t_0) \rangle \; , 
\label{eq:INI}
\end{equation}
which evolves in the course of one period~$T$ into the state
\begin{equation}
	| \psi(t_0 + T) \rangle = \sum_n a_n | u_n(t_0) \rangle 
	\exp(-\ri\varepsilon_n T/\hbar) \; . 
\end{equation}
Hence, for any $\varepsilon$ the absolute value of the return amplitude after 
one period~$T$ (``raT'') is given by
\begin{eqnarray}
	{\rm raT} & = & \Big| \exp(-\ri\varepsilon T/\hbar)
	\; \langle \psi(t_0 + T) | \psi(t_0) \rangle \, \Big|
\nonumber \\	& = &
	\left| \, \sum_n |a_n|^2 
	\exp\!\big(\ri[\varepsilon_n - \varepsilon]T/\hbar\big) 
	\, \right|	
\nonumber \\	& \ge &
	\sum_n | a_n |^2 
	\cos\!\big([\varepsilon_n - \varepsilon]T/\hbar\big)		 
\nonumber \\	& \ge &
	1 - \frac{T^2}{2\hbar^2} \sum_n | a_n |^2
	\big( \varepsilon_n - \varepsilon \big)^2 \; ,
\label{eq:RAT}
\end{eqnarray}
having used $\cos(x) \ge 1 - x^2/2$, and the normalization of the 
state~(\ref{eq:INI}). On the other hand, ``lifting'' that state to the
extended Hilbert space, thus considering  
\begin{equation}
	| \psi(t) \rangle = \sum_n a_n | u_n(t) \rangle 
\end{equation}
as an element of $L_2[0,T] \otimes {\mathcal H}$, one observes
\begin{equation}
	\sum_n | a_n |^2 
	\big( \varepsilon_n - \varepsilon \big)^2
	= F_\varepsilon \big[ | \psi \rangle \big] \; . 
\end{equation}
With the estimate~(\ref{eq:RAT}) holding for any~$\varepsilon$, this gives
\begin{equation}
	{\rm raT} \ge 1 - \min_\varepsilon \; 
	2\pi^2 \frac{F_\varepsilon}{(\hbar\omega)^2} \; .
\end{equation}
In particular, if $|\psi(t)\rangle$ is a Floquet function, one has
$\min_\varepsilon F_\varepsilon \big[ | \psi \rangle \big] = 0$, 
giving ${\rm raT} = 1$, as required. Hence, the value of the minimized 
functional $F_\varepsilon$ quantifies the failure of the variational solution 
to return to the initial state after one period, which may be taken as a 
measure of the ``quality'' of the approximate Floquet state obtained in this 
manner.

\section{Example}
\label{S_4}

Let us consider a one-dimensional harmonic oscillator which is subjected to
a monochromatic force with angular frequency~$\omega$, as described in the 
position representation by the Hamiltonian 
\begin{equation}
	H(t) = \frac{p^2}{2M} + \frac{1}{2} M \omega_0^2 x^2
	+ \lambda x \cos(\omega t) \; ,
\label{eq:HDO}
\end{equation}
where $M$ denotes the mass of the oscillator particle, $\omega_0$ is the
oscillator's angular frequency, and $\lambda$ specifies the amplitude of the
driving force. This is one of the few nontrivial Floquet systems which can be 
solved ana\-lytically~\cite{Husimi53,PopovPerelomov70,BreuerHolthaus89}, thus 
providing a benchmark test for the variational principle~(\ref{eq:FVP}).  

To begin with, let us asssume that the driving frequency~$\omega$ differs 
from the oscillator frequency~$\omega_0$, because the quasienergy spectrum 
of the forced oscillator~(\ref{eq:HDO}) becomes absolutely continuous in the 
resonant case $\omega = \omega_0$~\cite{HagedornEtAl86}. The construction
of the Floquet states then is based on the particular classical trajectory 
which shares the period $T = 2\pi/\omega$ of the driving force, that is, 
on the $T$-periodic solution to the classical equation of motion 
\begin{equation}
	\ddot{\xi} = -\omega_0^2 \xi - \frac{\lambda}{M} \cos(\omega t) \; ,
\end{equation}
which is
\begin{equation} 	  
	\xi(t) = \frac{\lambda}{M(\omega^2 - \omega_0^2)} \cos(\omega t) \; .
\label{eq:CTR}
\end{equation}
Denoting the familiar eigenfunctions of the unforced oscillator with 
energy eigenvalues $E_n = \hbar\omega_0(n + 1/2)$ by $\chi_n(x)$, where
$n = 0,1,2,\ldots$ is the usual oscillator quantum number, the desired 
Floquet states can be written as
\begin{eqnarray}
	\psi_n(x,t) & = & \chi_n\!\big( x - \xi(t) \big)
	\exp\!\left( \frac{\ri}{\hbar} M \dot{\xi}(t) \big( x - \xi(t) \right)
	\cdot  
\nonumber \\	& & \cdot		 
	\exp\!\left( -\frac{\ri}{\hbar} \! \left[
	E_n t - \int_0^t \! \rd \tau \, L(\tau) \right] \right) \; ,
\label{eq:FSO}
\end{eqnarray}
where
\begin{equation}
	L(t) = \frac{1}{2}M \dot{\xi}^2(t) - \frac{1}{2} M \omega_0^2 \xi^2(t)
	- \lambda \xi(t) \cos(\omega t)
\end{equation}	
is the classical Lagrangian of the system, evaluated along the 
trajectory~(\ref{eq:CTR}). Observing that the integral over this $T$-periodic
function~$L(t)$ contains a secular term which increases linearly with time,
and thus contributes to the respective quasienergy, the quasienergy spectrum
of the non-resonantly forced harmonic oscillator~(\ref{eq:HDO}) is deduced to 
read
\begin{eqnarray}
	\varepsilon_n & = & E_n - \frac{1}{T} \int_0^T \! \rd \tau \, L(\tau)
	\quad \bmod \hbar\omega
\nonumber \\	& = &
	\hbar\omega_0\!\left( n + \frac{1}{2} \right)
	+ \frac{\lambda^2}{4M(\omega^2 - \omega_0^2)}
	\quad \bmod \hbar\omega \; ,	
\label{eq:QES}
\end{eqnarray}
so that all states exhibit exactly the same ac Stark shift. This is a fairly 
unusual feature which reflects the integrability of the system~(\ref{eq:HDO}). 
Thus, apart from a phase factor the Floquet states~(\ref{eq:FSO}) are given by 
harmonic-oscillator eigenfunctions which follow the $T$-periodic oscillations 
of the classical trajectory~(\ref{eq:CTR}).

In order to explore whether these Floquet states are correctly recovered by 
the variational principle~(\ref{eq:FVP}), one may take   
\begin{equation}
	| \Psi \rangle = \sum_{n = \max(n_0-r,0)}^{n_0+r} \sum_{m=-r}^{r}
	c_{n,m} \, \re^{\ri m \omega t} \, | n \rangle 
\label{eq:ANS}
\end{equation}	 
as a natural general ansatz, with real coefficients $c_{n,m}$ to be used 
as variational parameters, and $\langle x | n \rangle = \chi_n(x)$. This 
ansatz now is employed for ``tracing'' the Floquet state which develops 
from the unperturbed oscillator ground state~$|0\rangle$ when the driving 
amplitude~$\lambda$ is gradually increased, while the driving 
frequency~$\omega$ is kept fixed. The procedure is as follows: 
For $\lambda = 0$ one has the exact solution $|\Psi\rangle = |0\rangle$ with 
$\varepsilon = E_0$, giving $F_\varepsilon \big[ | \Psi \rangle \big] = 0$. 
Then $\lambda$ is increased by a small amount $\delta\lambda$, and the 
variational state is seeded with the previous $| \Psi \rangle$. The variational
state then is propagated by one step in imaginary ``time'', allowing the state 
to relax towards the ground state of $(K - \varepsilon)^2$, and $\varepsilon$ 
is updated to the value $\lF K \rF$ resulting from the propagated state; this 
is repeated until the value of the functional $F_{\varepsilon}$ has numerically
converged to the accuracy specified. If, after convergence, the value of 
$F_{\varepsilon}/(\hbar\omega)^2$ equals zero within an acceptable tolerance, 
an approximate Floquet state has been found; if not, the searched Floquet state
is not contained in the selected variational space.

\begin{figure}[t]
\centering
\includegraphics[width=0.9\linewidth]{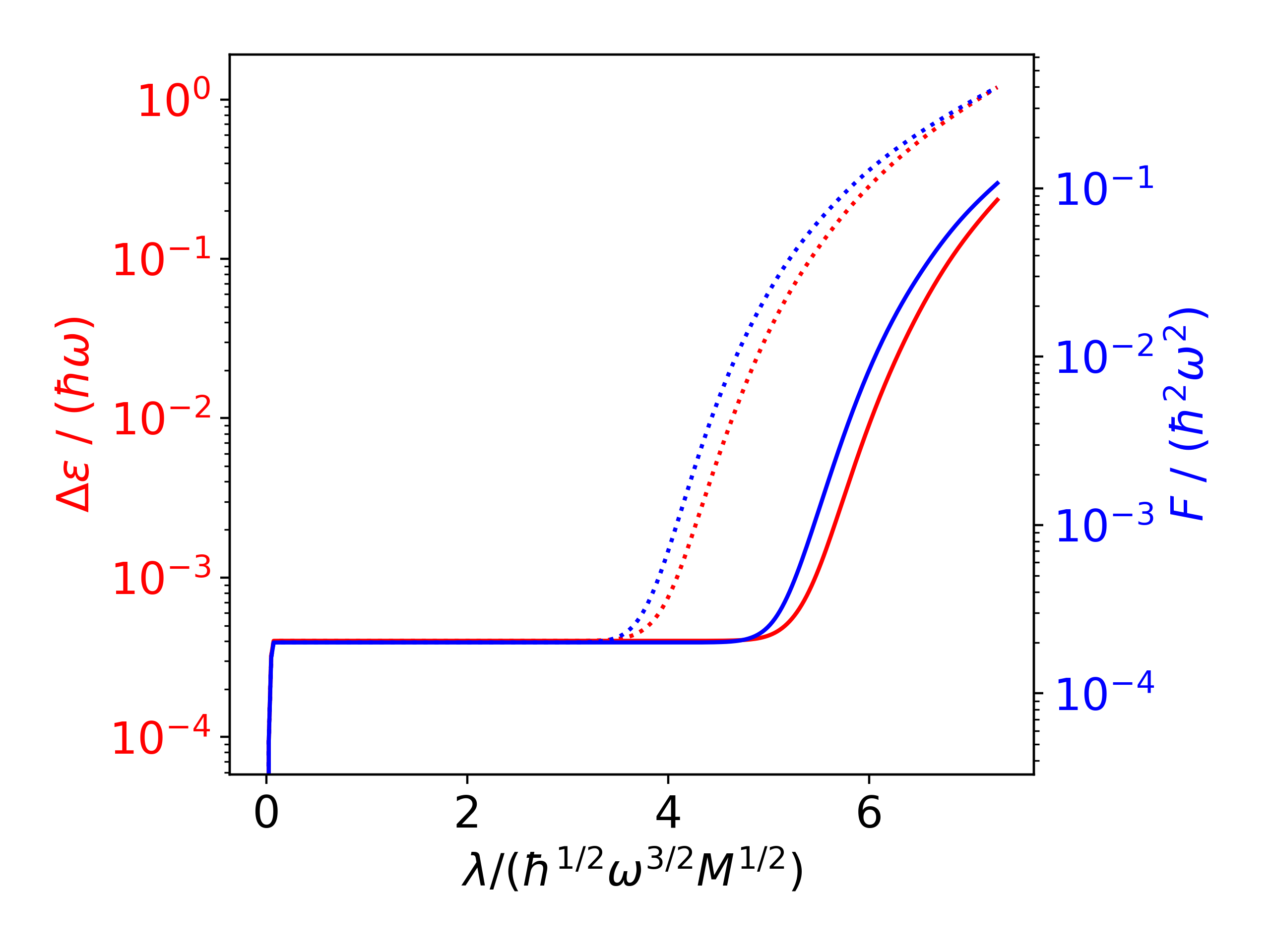}
\caption{Red: Difference $\Delta\varepsilon$ (absolute value) of the 
	variationally computed quasienergy $\varepsilon_0$ of the Floquet state
	which develops continuously from the ground state $n = 0$ of the 
	unforced oscillator, and the exact quasienergy~(\ref{eq:QES}), vs.\ 
	the scaled driving amplitude, as resulting for $\omega/\omega_0 = 2/3$
	from the ansatz~(\ref{eq:ANS}) with $n_0 = 0$ and $r = 20$ (dotted 
	lines) or $r = 30$ (full lines). Blue: Corresponding values of the 
	variational functional~(\ref{eq:FVP}) after minimization.}
\label{F_1}
\end{figure}

Figure~\ref{F_1} displays data computed according to this procedure for 
the driving frequency $\omega/\omega_0 = 2/3$, while the parameters~$n_0$ 
and~$r$ specifying the variational space have been chosen as $n_0 = 0$ and 
$r = 20$ (dotted lines) or $r = 30$ (full lines), respectively. Shown here 
are the absolute value $\Delta\varepsilon$ of the difference between the 
variationally obtained~$\varepsilon$ and the exact quasienergy obtained 
from Eq.~(\ref{eq:QES}) (red), together with the converged 
functional~$F_\varepsilon$ (blue) vs.\ the driving amplitude, 
all scaled to be dimensionless. Here the stepsize 
$\delta\lambda/\sqrt{\hbar M\omega^3} \approx 0.025$ has been employed;
the minimization procedure has been stopped when the absolute value of the
difference between the updated value of $F_\varepsilon/(\hbar\omega)^2$ 
and the previous one has become lower than $\Delta = 10^{-8}.$  
These results are extremely encouraging: Considering the data for $r = 20$ 
first, one observes a plateau at low driving amplitudes where the variational 
functional adopts an almost constant small value, 
$F_\varepsilon/(\hbar^2\omega^2) \approx 2 \cdot 10^{-4}$, while 
$\Delta\varepsilon/(\hbar\omega)$ likewise remains small, clearly signaling
that a good approximation to the exact Floquet state has been found. But
then, at $\lambda/\sqrt{\hbar M\omega^3} \approx 4$, the variational solution 
suddenly becomes inacceptable. This obervation can be understood with the help 
of a rough estimate: According to Eq.~(\ref{eq:FSO}), the exact Floquet state 
emanating from the unperturbed oscillator ground state is given by a Gaussian 
which sloshes along the classical trajectory~(\ref{eq:CTR}). On the other
hand, the basis states $\chi_n(x)$ comprising the variational space are
appreciably large in the classically allowed region only, that is, between the 
turning points~$x_{\rm tp}$ which limit the classical motion with energy~$E_n$ 
in the oscillator potential, $x_{\rm tp} = \pm \sqrt{2E_n/(M\omega_0^2)}$. 
In order to correctly represent the exact Floquet state within the variational 
space, the amplitude of the sloshing motion should be somewhat smaller than 
the largest of these turning points, implying     
\begin{equation}
	\left| \frac{\lambda}{M(\omega^2 - \omega_0^2)} \right| \le
	\sqrt{\frac{2E_r}{M\omega_0^2}} 
\end{equation}
and thus providing an upper bound on the driving amplitude that can be 
reasonably dealt with in a variational space made up from the lowest~$r+1$ 
oscillator functions, namely,  
\begin{equation}
	\frac{\lambda^2}{\hbar M\omega^3} \le
	\Big[ 1 - (\omega_0/\omega)^2 \Big]^2 \frac{\omega}{\omega_0}
	(2r + 1) \; .	
\label{eq:OME}	
\end{equation}
Inserting $\omega/\omega_0 = 2/3$, and $r = 20$, one finds
$\lambda/\sqrt{\hbar M\omega^3} \le 6.5$, which is in acceptable agreement 
with the behavior shown by the dotted lines in Fig.~\ref{F_1}, keeping in 
mind that the actual ``critical'' driving amplitude should be somewhat smaller
than the order-of-magnitude estimate~(\ref{eq:OME}).

Thus, when the variational space is enlarged, larger driving amplitudes
become admissible. For instance, when~$r$ is increased to $r = 30$, while
$\omega/\omega_0$ is kept fixed, the estimate~(\ref{eq:OME}) gives 
$\lambda/\sqrt{\hbar M\omega^3} \le 8.0$. That is, the range of manageable
driving amplitudes is increased by a factor of~$1.2$ in comparson with
the previous calculation, in fair agreement with the numerical data displayed 
in Fig~\ref{F_1}.

\begin{figure}[t]
\centering
\includegraphics[width=0.9\linewidth]{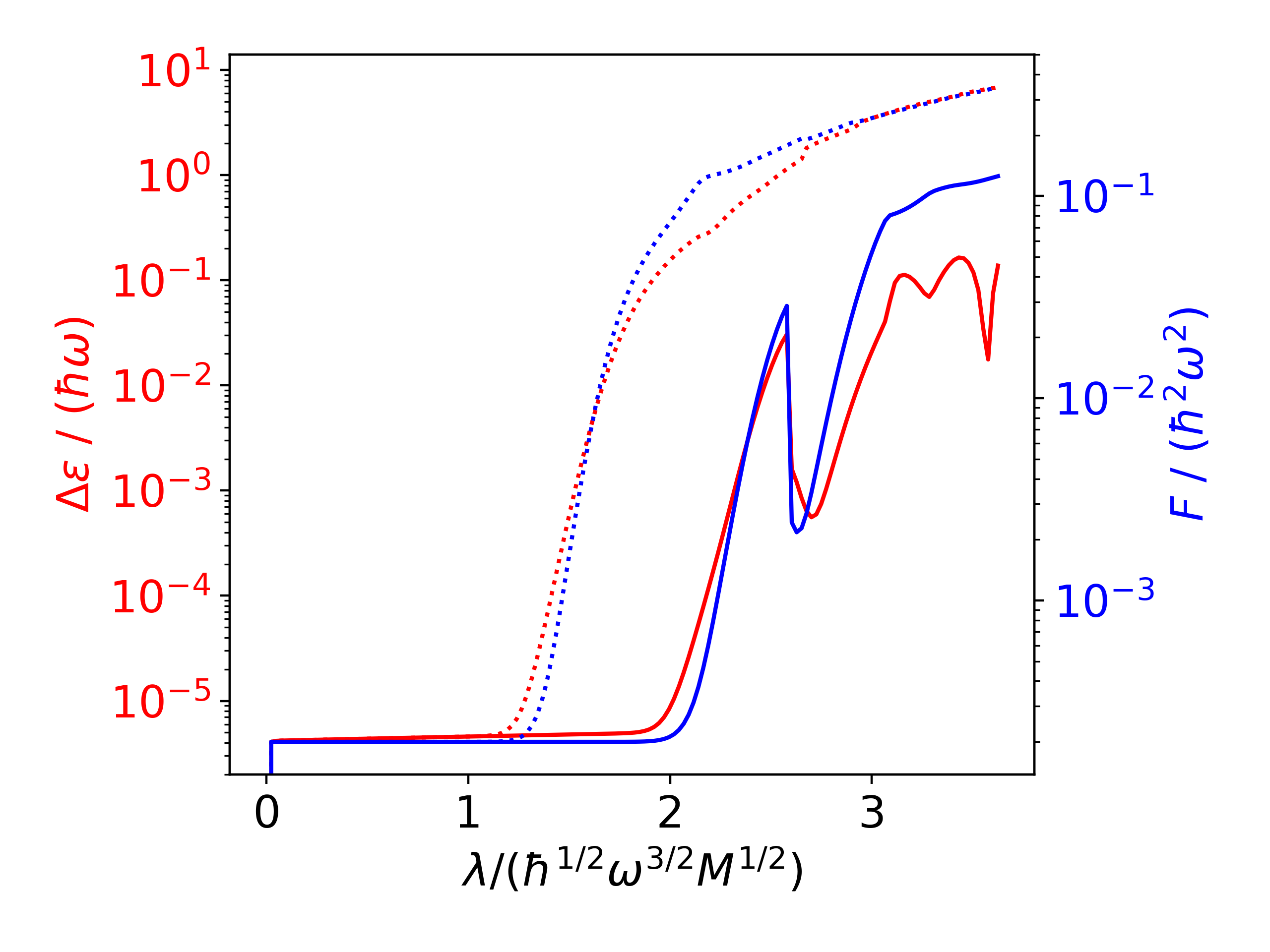}
\caption{As Fig.~\ref{F_1} for $n = n_0 = 50$, again with $r = 20$ 
	(dotted lines) and $r = 30$ (full lines).}
\label{F_2}
\end{figure}

The numerical strategy suggested in this work for following an individual
Floquet state through parameter space is not restricted to the Floquet state 
emanating from the ground state of the system in the absence of the drive, 
but applies the {\em any\/} state. In order to substantiate this claim, 
Fig.~\ref{F_2} depicts analogous numerical data obtained when tracing
the Floquet state originating from the unperturbed oscillator state $n = 50$,
again for $\omega/\omega_0 = 2/3$, computed in variational spaces spanned
by the ansatz states~(\ref{eq:ANS}) with $n_0 = 50$ and $r = 20$ (dotted lines)
or $r = 30$ (full lines). Once again, the results speak for themselves: The 
driving amplitudes above which the exact Floquet state is no longer adequately
represented by the respective variational ansatz are well discernible; below
these amplitues the variational principle~(\ref{eq:FVP}) provides excellent
approximations.

\section{Conclusion}
\label{S_5}

The customary computational strategies for determining the Floquet states of a 
periodically time-dependent quantum system rely either on Eq.~(\ref{eq:SEP}),
requiring the computation and diagonalization of the system's one-cycle
evolution operator~$U(T,0)$, or on the ``extended'' eigenvalue 
problem~(\ref{eq:EVP}), enforcing the use of a sufficiently large basis of
$L_2[0,T] \otimes {\mathcal H}$. While this may be no problem when dealing 
with periodically driven single-particle systems, say, it soon becomes 
impractical when investigating periodically driven many-body systems. For 
such high-dimensional systems the variational principle~(\ref{eq:FVP}) may 
unfold its full power, enabling one to compute individual Floquet states even 
for systems so large that the determination of the full quasienergy spectrum 
would be neither feasible nor even desirable. 

The particular model system that has been employed here for testing the
new variational principle, the linearly driven harmonic oscillator, certainly
is not typical from the Floquet point of view: Being explicitly integrable,
its quasienergies~(\ref{eq:QES}) and Floquet states~(\ref{eq:FSO}) can be  
labeled by the quantum number~$n$ of the harmonic-oscillator state to which
they are continuously connected when the driving amplitude goes to zero.
This is no longer the case for more generic systems, such as periodically 
forced anharmonic oscillators which possess a classial conterpart exhibiting 
chaotic dynamics. In such systems one encounters a quasienergy spectrum with
a dense net of anticrossings~\cite{Holthaus16,HoneEtAl09}, thwarting the notion 
of continuity. Nonetheless, it is surmised that the idea of ``tracing'' an 
individual Floquet state through para\-meter space will also work for such 
more realistic systems, helping one to identify those Floquet states which 
are most important for understanding a given system's experimentally observable
properties.

Finally, it needs to be stressed that the far-reaching progress made recently
in the areas of machine learning and hardware design will allow one to solve
variational problems even with very large numbers of variational parameters
in the near future. Therefore, it is anticipated that the combined use of 
the variational principle~(\ref{eq:FVP}), modern hardware, and intelligent
algorithms will enable one to investigate truly large Floquet systems which 
are way beyond the realm of the previously used standard numerical methods.

\begin{acknowledgments}
This work has been supported by the Deutsche For\-schungsgemeinschaft (DFG,
German Research Foundation) through Project No.~397122187. The author wishes 
to thank the members of the Research Unit FOR~2692 for many stimulating 
discussions. 
\end{acknowledgments}

\end{document}